\documentclass[aps,pra,showpacs,superscriptaddress,preprint]{revtex4}
\usepackage{amsmath}
\usepackage{epsf}
\usepackage{epsfig}

\catcode`ð=\active
 \defð{\u{g}}
 \catcode`Ð=\active
 \defÐ{\u{G}}
 \catcode`Ý=\active
 \defÝ{\. I}
 \catcode`ö=\active
 \defö{\"{o}}
 \catcode`Ö=\active
 \defÖ{\"O}
 \catcode`ü=\active
 \defü{\"{u}}
 \catcode`Ü=\active
 \defÜ{\"{U}}
 \catcode`Þ=\active
 \defÞ{\c{S}}
 \catcode`þ=\active
 \defþ{\c{s}}
 \catcode`ý=\active
 \defý{{\i}}
 \catcode`ç=\active
 \defç{\d{c}}
 \catcode`Ç=\active
 \defÇ{\d{C}}

\begin{document}

\title{Approximate Solution of the effective mass Klein-Gordon Equation
for the Hulthen Potential with any Angular Momentum}

\author{\small Altuð Arda}
\email[E-mail: ]{arda@hacettepe.edu.tr}\affiliation{Department of
Physics Education, Hacettepe University, 06800, Ankara,Turkey}
\author{Ramazan Sever}
\email[E-mail: ]{sever@metu.edu.tr}\affiliation{Department of
Physics, Middle East Technical University, 06800, Ankara,Turkey}
\date{\today}
\begin{abstract}
The radial part of the effective mass Klein-Gordon equation for the
Hulthen potential is solved by making an approximation to the
centrifugal potential. The Nikiforov-Uvarov method is used in the
calculations. Energy spectra and the corresponding eigenfunctions
are
computed. Results are also given for the case of constant mass.
\end{abstract}

\maketitle

\newpage

\section{Introduction}
There has been a continues interest on the solutions of the
Klein-Gordon (KG) equation and Dirac equation for some certain
potentials. The KG equation is solved by assuming that the scalar
potential equals or not equals to the vector potential, such as
Rosen-Morse-type potentials, Morse-like potential [1-8]. On the
other hand, it is also solved for the case of mixing of scalar and
vector potentials, such as vector-scalar Coulomb, kink-like
potentials, harmonic oscillator, Hartman potential, and
Hulthen-type potential [9-13]. Further, the Dirac equation is
solved for different types of potentials, such as harmonic, and
linear potentials [14], for an uniform magnetic field [15],
generalized asymmetrical Hartmann potentials [16]. Various methods
are used in the solutions, such as Nikiforov-Uvarov (NU) method,
by using the hypergeometric type equations, and separation of
variables [17-21].

In recent years, the effects of the position-dependent the mass on
the energy spectra and corresponding eigenvalues of the above
equations has been received a great attention [22, 23]. The exact
solutions of the above equations, and the Schrödinger equation in
the context of spatially dependent mass have been studied by many
authors [24-28]. Many different types of mass distributions have
been used in literature, such as polynomial mass function,
exponential, and hyperbolic mass distributions [30]. Our mass
function is similar to that used in Ref [24] by Dutra, and Almeida.
In the first example, the mass function, and potantial are taken as
In the present work, we prefer an exponential type  mass function to
fnd out the energy spectrum and corresponding wave functions of the
Hulthen potential. The Hulthen potential has applications in the
wide range of different areas such as nuclear, and particle physics,
atomic physics, solid state, and chemical physics [7].

In this study, we solve the radial KG equation for the Hulthen
potential by using NU-method within the framework of an
approximation to the centrifugal potential. The NU-method is based
on solving the second order equation by reducing to a generalized
equation of hypergeometric type [31].

The organization of this work is as follows. In Section II, we
solve the radial part of the KG equation by using an approximation
for the centrifugal term [33, 34], and compute energy eigenvalues
of the bound states and the corresponding eigenfunctions. We give
our conclusions in Section III.

\section{Nikiforov-Uvarov Method and Bound States}
The radial part of the KG equation reads [29]

\begin{eqnarray}
\left\{\,\frac{d^2}{dr^2}\,-\,\frac{\ell(\ell+1)}{r^2}\,-\,\frac{1}{\hbar^2c^2}\,
[m^2c^4-(E-V)^2]\right\}\phi(r)=0\,,\,\,\,(0 \leq r \leq
\infty)\,,
\end{eqnarray}
where $\ell$ is the angular-momentum quantum number, $E$ is the
energy of the particle, $c$ is the velocity of the light.

The Hulthen potential has the form [32]

\begin{eqnarray}
V(r)=\,-V_0\,\frac{e^{-\beta r}}{1-e^{-\beta r}}\,.
\end{eqnarray}
where $V_0$, and $\beta \equiv 1/a$ are constant parameters.

Eq. (1) can not be solved exactly because of the centrifugal
potential. So, one has to use an approximation for this term. This
approximation can be taken as [33, 34]

\begin{eqnarray}
\frac{\ell(\ell+1)}{r^2}\,\simeq\,\beta^2\ell(\ell+1)\,\frac{e^{-\beta
r}}{(1-e^{-\beta r})^2}\,,
\end{eqnarray}

There are very different mass-distributions are used in the
literature, such as an exponentially, and quadratically mass
functions [24], trigonometric mass-distributions, and mass
functions of the forms $m(r)=r^{\alpha}$, especially for
three-dimensional problems [26]. Here, we prefer to use the
following spatially dependent mass

\begin{eqnarray}
m(r)=m_0-\frac{m_1}{1-e^{-\beta r}}\,,\,\,\,(m_0 > m_1)\,.
\end{eqnarray}
which corresponds to a decay particle, and provides us an
approximate solution of the radial part of the effective KG
equation. $m_0$ and $m_1$ in this distribution are two arbitrary
positive parameters. The mass function of that form enable us to
check out the results in the limit of the constant mass. 2.  It is
well known that the Schrodinger equation (SE) should be reproduced
in the case of position-dependent mass, because of the
ordering-ambiguity problem between momentum, and mass operators in
kinetic energy term. The kinetic energy term should be written as
$p\,\frac{1}{2m(x)}\,p$. When the mass depends on coordinate, and it
can be seen that the operators no longer commute[see for details Ref
[10] by Gonul, and Ref [24]]. The ordering-ambiguity problem does
not appear in the case of KG equation.

Substituting Eqs. (2) and (4) into Eq. (1), and by using Eq. (3), we
get

\begin{eqnarray}
\Big\{\,\frac{d^2}{dr^2}\,-\beta^2\ell(\ell+1)\,\frac{e^{-\beta
r}}{(1-e^{-\beta
r})^2}\,+\,\frac{1}{\hbar^2c^2}\,\Big[\,E^2+2EV_0\,\,\frac{e^{-\beta
r}}{1-e^{-\beta r}}+V^2_0\,\,\frac{e^{-2\beta r}}{(1-e^{-\beta
r})^2}\nonumber\\-m^2_0c^4\,+\,\frac{2m_0m_1c^4}{1-e^{-\beta
r}}\,-\,\frac{m^2_1c^4}{(1-e^{-\beta r})^2}\Big]\Big\}\phi(r)=0\,.
\end{eqnarray}
By using the transformation $z=1-e^{-\beta r}\,\,(0 \leq z \leq
1)$, we obtain

\begin{eqnarray}
\frac{d^2\phi(z)}{dz^2}\,-\,\frac{z}{z(1-z)}\,\frac{d\phi(z)}{dz}\,+\,\frac{1}{[z(1-z)]^2}
\left[-a_1^2z^2-a_2^2z-a_3^2\right]\phi(z)=0\,,
\end{eqnarray}
where

\begin{eqnarray}
-a^2_1&=&Q^2(E^2-2EV_0+V^2_0-m^2_0c^4)\,,\nonumber\\
-a^2_2&=&Q^2(2m_0m_1c^4+2EV_0-2V^2_0)\,+\,\ell(\ell+1)\,,\nonumber\\
-a^2_3&=&Q^2(V^2_0-m^2_1c^4)\,-\,\ell(\ell+1)\,.
\end{eqnarray}
and $Q^2=1/\beta^2\hbar^2c^2$. Now to apply the NU-method [31], we
rewrite Eq. (6) in the following form

\begin{eqnarray}
\phi^{\prime\prime}(z)+\,\frac{\tilde{\tau}(z)}{\sigma(z)}\,\phi^{\prime}(z)
+\,\frac{\tilde{\sigma}(z)}{\sigma^2(z)}\,\phi(z)=0,
\end{eqnarray}
where $\sigma(z)$ and $\tilde{\sigma}(z)$ are polynomials with
second-degree, at most, and $\tilde{\tau}(z)$ is a polynomial with
first-degree. We define a transformation for the total wave
function as

\begin{eqnarray}
\phi(z)=\xi(z)\psi(z).
\end{eqnarray}
Thus Eq. (8) is reduced to a hypergeometric type equation

\begin{eqnarray}
\sigma(z)\psi^{\prime\prime}(z)+\tau(z)\psi^{\prime}(z)+\lambda\psi(z)=0.
\end{eqnarray}
We also define the new eigenvalue for the Eq. (8) as

\begin{eqnarray}
\lambda&=&\lambda_n=-n\tau^{\prime}-\,\frac{n(n-1)}{2}\,\sigma^{\prime\prime}\,,
(n=0, 1, 2, \ldots)
\end{eqnarray}
where

\begin{eqnarray}
\tau(z)&=&\tilde{\tau}(z)+2\pi(z).
\end{eqnarray}
The derivative of $\tau(z)$ must be negative. $\lambda(\lambda_n)$
is obtained from a particular solution of the polynomial
$\psi_n(z)$ with the degree of $n$. $\psi_n(z)$ is the
hypergeometric type function whose solutions are given by [31]

\begin{eqnarray}
\psi_n(z)=
\,\frac{b_n}{\rho(z)}\,\frac{d^n}{dz^n}[\sigma^n(z)\rho(z)],
\end{eqnarray}
where the weight function $\rho(z)$ satisfies the equation

\begin{eqnarray}
\frac{d}{dz}[\sigma(z)\rho(z)]=\tau(z)\rho(z).
\end{eqnarray}
On the other hand, the function $\xi(z)$ satisfies the relation

\begin{eqnarray}
\xi^{\prime}(z)/\xi(z)=\pi(z)/\sigma(z).
\end{eqnarray}
Comparing Eq. (6) with Eq. (8), we have

\begin{eqnarray}
\tilde{\tau}(z)=-z\,,\,\,\,\,\,\sigma(z)=z(1-z)\,,\,\,\,\,\,
\tilde{\sigma}(z)=-a_1^2z^2-a_2^2z-a_3^2
\end{eqnarray}
The $\pi(z)$ has the form

\begin{eqnarray}
\pi(z)=\,\frac{\sigma^{\prime}(z)-\tilde{\tau}(z)}{2}\,
\pm\,\sqrt{(\frac{\sigma^{\prime}(z)-\tilde{\tau}(z)}{2})^2-\tilde{\sigma}(z)+k\sigma(z)}\,,
\end{eqnarray}
or, explicitly

\begin{eqnarray}
\pi(z)=\,\frac{1}{2}\,(1-z)\,\mp\sqrt{(\,\frac{1}{4}\,+a_1^2-k)z^2+(a_2^2+k\,\frac{1}{2}\,)z+a_3^2+\,\frac{1}{4}\,}\,,
\end{eqnarray}
The constant $k$ is determined by imposing a condition such that
the discriminant under the square root should be zero. The roots
of $k$ are $k_{1,2}=-a_2^2-2a_3^2\mp A\sqrt{1+4a^2_3}$, where
$A=\sqrt{a_3^2+a_2^2+a_1^2}$. Substituting these values into
Eq.(18), we get for $\pi(z)$

\begin{eqnarray}
\pi(z)(k \rightarrow k_1)=\,\frac{1}{2}\,(1-z)
\mp\Big[\Big(A-\sqrt{\,\frac{1}{4}\,+a^2_3}\,\Big)z+\sqrt{\,\frac{1}{4}\,+a^2_3}\,\Big]\,,
\end{eqnarray}
and

\begin{eqnarray}
\pi(z)(k \rightarrow k_2)=\,\frac{1}{2}\,(1-z)
\mp\Big[\Big(A+\sqrt{\,\frac{1}{4}\,+a^2_3}\,\Big)z-\sqrt{\,\frac{1}{4}\,+a^2_3}\,\Big]\,,
\end{eqnarray}
Now we calculate the polynomial $\tau(z)$  from $\pi(z)$ such that
its derivative with respect to $z$ must be negative. Thus we take
the first choice

\begin{eqnarray}
\tau(z)=1-2\sqrt{\,\frac{1}{4}\,+a^2_3}\,-2\Big(\,A-\sqrt{\,\frac{1}{4}\,+a^2_3}\,+1\Big)z.
\end{eqnarray}
The constant $\lambda=k+\pi^{\prime}(z)$ becomes

\begin{eqnarray}
\lambda=-a_2^2-2a_3^2+A\sqrt{1+4a^2_3}\,-\,\frac{1}{2}\,-\Big(\,A-\sqrt{\,\frac{1}{4}\,+a^2_3}\,\Big)\,,
\end{eqnarray}
and Eq. (11) gives us

\begin{eqnarray}
\lambda_n=2n\Big(\,A-\sqrt{\,\frac{1}{4}\,+a^2_3}\,+1\Big)+n^2-n\,.
\end{eqnarray}
Substituting the values of the parameters given by Eq. (7), and
setting $\lambda=\lambda_n$, one can find the energy eigenvalues
as

\begin{eqnarray}
E_{n\ell}&=&\,\frac{V_0}{2}\,+\,\frac{1}{4Q^2(N^2+4Q^2V^2_0)}\Big\{8Q^3m_1c^4(m_1-2m_0)\nonumber\\&\mp&
N\Big[16Q^4(V^2_0-m^2_1c^4)(m^2_1c^4-4m_0m_1c^4+4m^2_0c^4-V^2_0)\nonumber\\
&+&8Q^2N^2(2m^2_0c^4-2m_0m_1c^4+m^2_1c^4-V^2_0)-N^4\Big]^{1/2}\Big\}\,,
\end{eqnarray}
where

\begin{eqnarray}
N=(2n+1)+\sqrt{1+4a^2_3\,}\,.
\end{eqnarray}
We see that the energy levels for particles and antiparticles are
symmetric, and the ground state energy is different from zero.

We also get the energy eigenfunctions for the constant mass case for
$s$-states

\begin{eqnarray}
E^{m_1=0}_{n,\ell=0}=\,\frac{V_0}{2}\,\pm\,N'\sqrt{\,\frac{m^2_0c^4}{4Q^2V^2_0+N'^2}\,-\,\frac{1}{16Q^2}}\,,
\end{eqnarray}
where

\begin{eqnarray}
N'=(2n+1)+\sqrt{1-4Q^2V^2_0\,}\,.
\end{eqnarray}
It is the same result with those in literature for $q=1$ [21].

Now let us find the eigenfunctions. We first compute the weight
function from Eqs. (12) and (14)

\begin{eqnarray}
\rho(z)=z^{\sqrt{1+4a^2_3\,}}\,(1-z)^{2A}\,,
\end{eqnarray}
and the wave functions become

\begin{eqnarray}
\psi_n(z)=\,\frac{b_n}{z^{\sqrt{1+4a^2_3\,}}\,(1-z)^{2A}}\,\frac{d^n}{dz^n}\,\left[
\,z^{n+\sqrt{1+4a^2_3\,}}\,(1-z)^{n+2A}\right]\,.
\end{eqnarray}
where $b_n$ is a normalization constant. The polynomial solutions
can be written in terms of the Jacobi polynomials [35]

\begin{eqnarray}
\psi_n(z)=b_n\,P_n^{(\sqrt{1+4a^2_3},\,
2A\,)}\,(1-2z)\,,\,\,\,\,\,2A>-1\,,\,\,\,\,\,\sqrt{1+4a^2_3\,}>-1\,.
\end{eqnarray}
On the other hand, the other part of the wave function is obtained
from the Eq. (15) as

\begin{eqnarray}
\xi(z)=z^{\,\frac{1}{2}\,\big[1+\sqrt{1+4a^2_3\,}\big]}\,(1-z)^{A}\,.
\end{eqnarray}
Thus, the total eigenfunctions take

\begin{eqnarray}
\phi_n(z)=b'_n\,(1-z)^{A}\,z^{\,\frac{1}{2}\,\big[1+\sqrt{1+4a^2_3\,}\big]}\,P_n^{(\sqrt{1+4a^2_3\,},\,2A\,)}\,(1-2z)\,.
\end{eqnarray}
where $b'_n$ is the new normalization constant.

Total wave function for the constant mass case for any $\ell$-state
takes

\begin{eqnarray}
\phi^{m_1=0}_{n\ell}(z)=b''_n\,(1-z)^{A'}\,z^{\,\frac{1}{2}\,\big[1+\sqrt{1+4a'^2_3\,}\big]}\,
P_n^{(\sqrt{1+4a'^2_3\,},\,2A'\,)}\,(1-2z)\,.
\end{eqnarray}
It is consistent with the results obtained in the literature [21].
The parameters for this case are given by
$A'=\sqrt{a_3^{'2}+a_2^{'2}+a_1^{'2}}$, and

\begin{eqnarray}
a_1^{'2}&=&a^2_1\,,\nonumber\\
a_2^{'2}&=&2Q^2V_0(V_0-E)-\ell(\ell+1)\,,\nonumber\\
a_3^{'2}&=&-Q^2V^2_0+\ell(\ell+1)\,.
\end{eqnarray}

\section{Conclusion}
We have obtained approximate solution of the radial part of the
effective mass KG equation for the Hulthen potential in the
framework of an approximation to the centrifugal term for any
$\ell$ values. We have obtained the energy spectra and the
corresponding radial part of the wave functions by applying the
NU-method. We have found a real energy spectra for the Hulthen
potential in the PDM case. To check our results, we have also
calculated the energy eigenvalues and eigenfunctions of the
particle and antiparticles for the case of constant mass limit for
$s$-waves, and seen that the results are consistent with those in
the literature [21].

\section{Acknowledgments}
This research was partially supported by the Scientific and
Technical Research Council of Turkey.


\newpage

\begin{thebibliography}{99}

\bibitem{ref1} Q.~W.-Chao, Chin. Phys. {\bf 11}, 757
(2002).

\bibitem{ref2} Q.~W.-Chao, Chin. Phys. {\bf 12}, 136
(2003).

\bibitem{ref3} L.-Z.~Yi, Y.-F.~Diao, J.-Y.~Liu, and C.-S.~Jia, Phys. Lett. A {\bf
333}, 212 (2004).

\bibitem{ref4} X.-Q.~Zhao, et al., Phys. Lett. A {\bf
337}, 189 (2005).

\bibitem{ref5} A.~S~.~de~Castro, [arXiv:~hep-th/0409216].

\bibitem{ref6} W.-C.~Qiang, R.-S.~Zhou, and Y.~Gao, Phys. Lett. A {\bf
371}, 201 (2007); S.~Dong, J.~Garcia-Ravelo, and S.~Dong, Phys.
Scr. {\bf 76}, 393 (2007); S.~M.~Ikhdair, and R.~Sever, J. Math.
Chem. {\bf 42}, 461 (2007).

\bibitem{ref7} N.~Saad, [arXiv:~math-ph/0709.4014].

\bibitem{ref8} C.-Y.~Chen, D.-S.~Sun, and F.~L.~Lu, Phys. Lett. A {\bf 370},
219 (2007).

\bibitem{ref9} L.~Chetouani, et al., Physica A {\bf 234}, 529 (1996).

\bibitem{ref10} F.~Dominguez-Adame, Phys. Lett. A {\bf 136}, 175
(1989); S.~M.~Ikhdair, and R.~Sever, J. Math. Chem. {\bf 42}, 461
(2007); B.~Gonul, Chin. Phys. Lett. {\bf 23}, 2640 (2006).

\bibitem{ref11} A.~S.~de~Castro, [arXiv:~hep-th/0511010].

\bibitem{ref12} A.~S.~de~Castro, [arXiv:~hep-th/0507218].

\bibitem{ref13} A.~D.~Alhaidari, H.~Bahlouili, and A.~Al-Hasan, Phys. Lett. A {\bf 349},
97 (2006).

\bibitem{ref14} R.~Giachetti, and E.~Sorace, [arXiv:~hep-th/0706.0127].

\bibitem{ref15} K.~Bhattacharya, [arXiv:~hep-th/0705.4275].

\bibitem{ref16} A.~S.~Dutra, and M.~B.~Hott, [arXiv:~quant-ph/0705.3447].

\bibitem{ref17} V.~M.~Villalba, and C.~Rojas, Int. J. Mod. Phys. A {\bf 21}, 313 (2006), [arXiv:~hep-th/0508040].

\bibitem{ref18} L.~A.~Gonzalez, and V.~M.~Villalba, Mod. Phys. Lett. A {\bf 20}, 2245 (2005), [arXiv:~math-ph/0507065].

\bibitem{ref19} V.~M.~Villalba, and E.~I.~Catala, J. Math. Phys. {\bf 43}, 4909 (2002), [arXiv:~gr-qc/0208017].

\bibitem{ref20} S.~M.~Ikhdair, and R.~Sever, Ann. Phys. (Leipzig){\bf 16}, 218 (2002), [arXiv:~quant-ph/0610183].

\bibitem{ref21} H.~Eðrifes, and R.~Sever, Int. J. Theo. Phys. {\bf 46}, 935 (2007), [arXiv:~quant-ph/0609231];
M.~Þimþek and H.~Eðrifes, J. Phys. A {\bf 37}, 4379 (2004)
[arXiv:~quant-ph/0211025].

\bibitem{ref22} T.~Gora and F.~Williams, Phys. Rev. {\bf 177}, 11979 (1969).

\bibitem{ref23} O.~von~Roos, Phys. Rev. B {\bf 27}, 7547 (1983).

\bibitem{ref24} A.~S.~Dutra, and C.~A.~S.~Almeida, Phys. Lett. A {\bf 275}, 25
(2000).

\bibitem{ref25} R.~Sever, and C.~Tezcan,
[arXiv:~quant-ph/0712.0268]; C.~Tezcan, and R.~Sever, J. Math.
Chem. {\bf 42}, 387 (2007), [arXiv:~quant-ph/0604041].

\bibitem{ref26} A.~D.~Alhaidari, Phys. Lett. A {\bf 322}, 72 (2004); A.~D.~Alhaidari, Phys. Rev. A {\bf 66}, 042116 (2002).

\bibitem{ref27} R.~Chen, [arXiv:~physics.gen-ph/0706.4147].

\bibitem{ref28} O.~Mustafa, and S.~H.~Mazharimousavi, [arXiv:~quant-ph/0611149].

\bibitem{ref29} M.~M.~Panja, R.~Dutt, and Y.~P.~Varshni, Phys. Rev. A {\bf 42}, 106 (1990);
M.~M.~Panja, and R.~Dutt, Phys. Rev. A {\bf 38}, 3937 (1998);
M.~M.~Panja, M.~Bag, R.~Dutt, and Y.~P.~Varshni, Phys. Rev. A {\bf
45}, 1523 (1992).

\bibitem{ref30} J.~G.-Xing, et al., [arXiv:~quant-ph/0707.3259].

\bibitem{ref31} A.~F.~Nikiforov, and V.~B.~Uvarov, \textit{Special Functions of
Mathematical Physics }, (Birkh\"{a}user, Basel, 1988).

\bibitem{ref32} S.~Flügge, \textit{Practical Quantum Mechanics I }, (Springer-Verlag, 1971).

\bibitem{ref33} R.~L.~Greene, and C.~Aldrich, Phys. Rev. A {\bf 14}, 2363 (1976).

\bibitem{ref34} W.-C.~Qiang, and S.~H.~Dong, Phys. Lett. A {\bf 368}, 13 (2007).

\bibitem{ref35} C.~W.~Wong, \textit{Introduction to Mathematical Physics-Methods
and Concepts }, (Oxford University Press, 1991).

\bibitem{ref36} G.~Szegö, \textit{Orthogonal Polynomials }, (Providence, RI: Amer.
Math. Soc., 1988).
\end{thebibliography}
\end{document}